\PassOptionsToPackage{numbers,sort&compress}{natbib}
\documentclass[aps,prab,reprint,superscriptaddress,nofootinbib,amsmath,amssymb,floatfix]{revtex4-2}

\usepackage{graphicx}
\usepackage{hyperref}
\usepackage{url}

\begin{document}

\title{Wake Perturbations in Laser- and Beam-Driven Plasma Wakefield Accelerators: A Symmetry-Based Multipole Classification}

\author{Andrei C. Berceanu}
\email{andrei.berceanu@eli-np.ro}
\affiliation{Extreme Light Infrastructure -- Nuclear Physics (ELI-NP) and Horia Hulubei National Institute for R\&D in Physics and Nuclear Engineering (IFIN-HH), 30 Reactorului Street, 077125 M\u{a}gurele, Romania}

\author{Alessio Del Dotto}
\affiliation{INFN -- Laboratori Nazionali di Frascati, Via Enrico Fermi 54, 00044 Frascati (RM), Italy}

\date{\today}

\begin{abstract}
We review beam-quality physics in laser-driven (LWFA) and beam-driven (PWFA) plasma wakefield accelerators through the symmetry group of the idealised blowout wake --- axisymmetry $\mathrm{SO}(2)_\phi$, adiabatic longitudinal translation, and propagation-direction parity. Transverse perturbations of the wake are classified by an integer azimuthal multipole order $m$ labelling the irreducible representations of $\mathrm{SO}(2)_\phi$, with the lowest beam-quality observables coupling at a specific multipole: the bunch centroid at $m{=}1$, cross-plane emittance coupling at $m{=}2$. A symplectic analogy relates transverse matching to longitudinal beam loading. Several phenomena common to LWFA and PWFA --- hose instabilities, pulse-front-tilt jitter, spot-asymmetry emittance growth, polarisation-dependent centroid motion, resonant cross-plane mixing --- populate the two lowest non-trivial $m$-channels and admit a unified discussion. The positron-witness problem reorganises in the same language: each known mitigation abandons one specific feature of the uniform-density blowout, drawn from a finite set. The classification also raises the possibility of an $m{=}3$ response channel whose magnitude remains open. We note the connection to symmetry-equivariant Bayesian optimisation of plasma accelerators.
\end{abstract}

\maketitle

\section{Introduction}
\label{sec:introduction}

Laser-driven (LWFA) and beam-driven (PWFA) plasma wakefield accelerators deliver GV/m accelerating gradients~\citep{tajima_laser_1979,esarey_physics_2009}, three orders of magnitude beyond conventional RF machines. The residual beam-quality requirements --- normalised emittance at the 100~nm level, sub-per-cent energy spread, pointing stability, per-stage emittance preservation in many-stage collider designs~\citep{albert_roadmap_2021,adolphsen_european_2022,gessner_design_2025} --- determine whether these devices can serve as useful FEL drivers, injectors for TeV-class linear colliders, or compact radiation sources. We review the beam-quality literature under a non-standard organising principle. The idealised blowout wake has a continuous symmetry group --- rotation about the driver axis, adiabatic translation in the co-moving coordinate, a discrete propagation-direction parity --- and the conserved quantities and irreducible representations of that group dictate which perturbations of the wake can couple to a given transverse moment of the accelerated bunch~\citep{arnold_mathematical_1989,schwichtenberg_physics_2015}.

Symmetry is the standard organising principle of conventional accelerator physics~\citep{courant_theory_1958,chao_physics_1993,reiser_theory_2008}, of quantum-field-theory spectra~\citep{schwichtenberg_physics_2015}, and of condensed-matter band structure~\citep{dresselhaus_group_2008}, but it is not usually the lens for plasma-acceleration literature. Existing reviews~\citep{esarey_physics_2009,malka_laser_2011,corde_femtosecond_2013,albert_roadmap_2021} organise the material along natural experimental lines --- by injection scheme, by driver type, by regime, while the organisation adopted here is complementary. A small number of symmetries of the wake and of the single-particle motion inside it project each degradation channel onto a definite azimuthal harmonic. Hosing~\citep{huang_hosing_2007,nechaeva_hosing_2024}, pointing jitter from pulse-front tilt~\citep{seidel_polarization_2022}, spot-asymmetry-induced emittance growth~\citep{moulanier_modeling_2023,manwani_analysis_2024}, polarisation-dependent centroid motion~\citep{seidel_polarization_2022}, and the resonant cross-plane emittance mixing in PWFA~\citep{diederichs_resonant_2025} appear in this organisation as five projections of the same multipole expansion onto the two lowest non-trivial harmonics.

The remainder of the paper is organised as follows. Section~\ref{sec:framework} introduces the symmetry group of the idealised blowout wake and the associated multipole expansion. Sections~\ref{sec:lwfa} and~\ref{sec:pwfa} apply the resulting selection rules to the LWFA and PWFA beam-quality literature, respectively. Section~\ref{sec:emittance} treats emittance preservation as a symplectic problem, covering matching, beam loading, and the azimuthal channels through which growth proceeds. Section~\ref{sec:conclusions} discusses open questions and the connection to symmetry-equivariant Bayesian optimisation.

\section{Symmetry framework for wakefield accelerators}
\label{sec:framework}

\subsection{The symmetry group of the idealised blowout wake}
\label{subsec:symmetry-group}

In the blowout regime~\citep{pukhov_laser_2002} the driver expels nearly all plasma electrons from a spheroidal cavity behind it, leaving a pure-ion channel whose transverse focusing force is linear in the displacement from axis. Under the quasi-static approximation, the cavity is stationary in the co-moving frame $\xi = z - ct$ and the fields depend only on $\xi$ and the cylindrical radius $r$. The idealised wake therefore has the symmetry group
\begin{equation}
\label{eq:symmetry-group}
G \;=\; \mathrm{SO}(2)_{\phi} \;\times\; T_{\xi}^{\rm slow} \;\times\; P_{z\leftrightarrow -z}\,,
\end{equation}
where $\mathrm{SO}(2)_{\phi}$ is rotation about the driver axis, $T_{\xi}^{\rm slow}$ is adiabatic translation in $\xi$, and $P_{z\leftrightarrow -z}$ is the discrete parity that takes forward propagation into backward propagation and sends ($e^-,e^+$) witnesses into one another. $T_{\xi}^{\rm slow}$ is exact for a rigid driver in uniform plasma and a slow-evolving adiabatic symmetry otherwise; in practice driver etching, self-steepening, and depletion~\citep{esarey_physics_2009} break it on the dephasing or depletion scale, and $T_{\xi}^{\rm slow}$-conservation statements in what follows apply rigorously only over propagation lengths short compared to the driver-evolution timescale.

The conserved quantities implied by~\eqref{eq:symmetry-group} structure the multipole classification developed in this section. The transverse and longitudinal forces on a relativistic witness derive from a single scalar pseudopotential $\Psi(r,\phi,\xi) \equiv \phi - A_z$~\citep{lu_nonlinear_2006,esarey_physics_2009}, with $\vec{F}_\perp = -\nabla_\perp \Psi$ and $F_z = -\partial_\xi \Psi$. $\mathrm{SO}(2)_{\phi}$ conservation of the canonical angular momentum $L_z$ forces $\Psi$ to decompose as a pure multipole series in cylindrical coordinates $(r,\phi,\xi)$,
\begin{equation}
\label{eq:multipole}
\Psi(r,\phi,\xi) \;=\; \sum_{m=0}^{\infty} \bigl[\,\Psi_m^{c}(r,\xi)\cos m\phi + \Psi_m^{s}(r,\xi)\sin m\phi\,\bigr]\,,
\end{equation}
where $\Psi_m^{c}(r,\xi)$ and $\Psi_m^{s}(r,\xi)$ are the cosine and sine multipole amplitudes at azimuthal order $m$. In the ideal axisymmetric cavity only the $m{=}0$ amplitude is nonzero, and all higher-$m$ amplitudes vanish. Any departure from this ideal --- driver offset, laser polarisation, ion motion, asymmetric beam loading~\citep{tzoufras_beam_2008} --- populates one or more $m{\geq}1$ harmonics of Eq.~\eqref{eq:multipole} and only those harmonics.

$T_{\xi}^{\rm slow}$ conservation provides an approximate Hamiltonian $H_{\xi}(\xi, p_\xi)$ governing longitudinal single-particle motion, including the beam-loading response of the wake~\citep{tzoufras_beam_2008}. The discrete parity $P_{z\leftrightarrow -z}$ underlies the electron--positron asymmetry of the plasma response addressed in Sec.~\ref{subsec:positron}: a positron sees the same $m{=}0$ cavity as an electron but with the opposite sign of transverse force. The asymmetry traces to the mass-ratio asymmetry of the plasma background --- electrons respond on the wake timescale while ions remain effectively fixed --- so a charge-conjugation operation applied to the witness alone does not symmetrise the response, even though the underlying QED Lagrangian is invariant under $e^- \leftrightarrow e^+$.

Equation~\eqref{eq:multipole} decomposes the wake fields into irreducible representations (irreps) of $\mathrm{SO}(2)_\phi$. The multipole index $m$ is the eigenvalue of the rotation generator $L_z = -i\,\partial_\phi$; fields of different $m$ belong to distinct irreps and are mutually orthogonal. Any $\mathrm{SO}(2)_\phi$-invariant linear response is diagonal in the $m$-basis, coupling only same-$m$ representations. The $m$-label we use throughout this review is therefore the minimal index that labels the irreducible content of any wake perturbation, and beam-quality channels that act through different $m$-irreps are decoupled at leading order by representation theory alone. This label is gauge-invariant, mechanism-independent, and fixed by the symmetry group of the wake rather than by any particular dynamical equation.

\begin{figure}[!htbp]
  \centering
  \includegraphics[width=\linewidth]{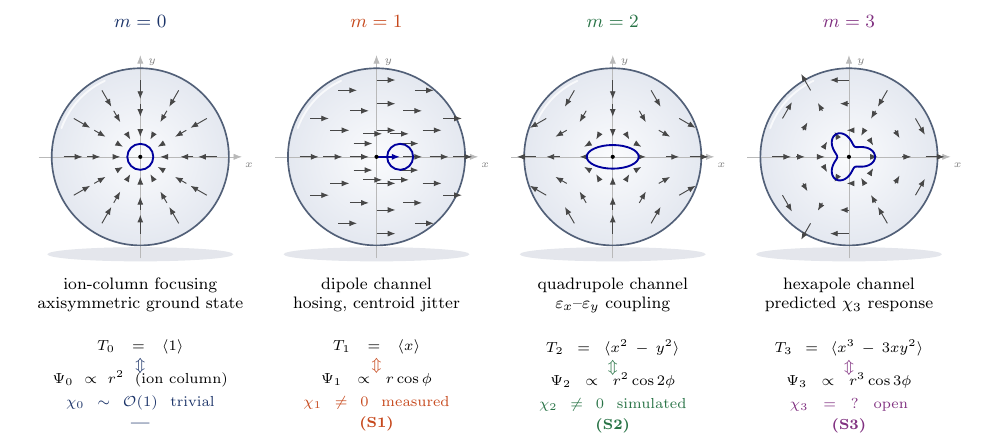}
  \caption{Schematic of the four lowest azimuthal-mode channels in the blowout wake [Eq.~\eqref{eq:multipole}], $m \in \{0,1,2,3\}$. Each column shows the transverse force field $\vec{F}_\perp = -\nabla\Psi_m$ for one channel (arrows) and the matched witness moment $T_m$ (blue lobes), with the propagation direction $\hat{z}$ normal to the page. Below each panel the pairing $T_m \leftrightarrow \Psi_m$, the response coefficient $\chi_m$, and the corresponding selection-rule tag (S1)/(S2)/(S3) are listed; see the surrounding text for the physical content of each channel.}
  \label{fig:azimuthal-modes}
\end{figure}

Figure~\ref{fig:azimuthal-modes} summarises the four lowest channels. Each pairs the multipole amplitude $\Psi_m^{c,s}$ of Eq.~\eqref{eq:multipole} with a matched witness moment $T_m$ that transforms as the same irrep --- $T_1 = \langle x \rangle$ is the bunch centroid, $T_2 = \langle x^2 - y^2 \rangle$ the quadrupole moment, $T_3 = \langle x^3 - 3xy^2 \rangle$ the hexapole moment --- and a response coefficient $\chi_m$ characterising how strongly the wake's $m$-th harmonic responds to a witness moment of the same order, with $\chi_m$ set by the plasma dynamics rather than by representation theory.

The four channels carry distinct physical content. The $m{=}0$ amplitude $\Psi_0^c = \tfrac14 k_p^2 r^2$ is the parabolic pseudopotential of the unperturbed cavity, generating the linear focusing $\vec{F}_\perp = -\tfrac12 k_p^2 \vec{r}$ from the uniform positive ion background ($k_p$ is the plasma wavenumber); it acts on the bunch rather than being driven by it. $m{=}1$ is populated whenever the bunch centroid leaves the cavity axis; the resulting dipole wake feeds back on the centroid and underlies the hose instability~\citep{huang_hosing_2007}. $m{=}2$ vanishes by axisymmetry in the ideal blowout, so $\chi_2 \neq 0$ requires contributions beyond the linear response, such as relativistic ion motion or beam-induced ionisation~\citep{diederichs_resonant_2025}. $m{=}3$ is allowed by representation theory; we discuss its status in Sec.~\ref{sec:conclusions}.

\subsection{Selection rules from \texorpdfstring{$\mathrm{SO}(2)$}{SO(2)} representation theory}
\label{subsec:selection-rules}

Transverse bunch moments decompose into irreducible representations of $\mathrm{SO}(2)_\phi$ labelled by an azimuthal index $m$, and each irreducible component is fixed by its transformation law under rotations of $\phi$. Any $\mathrm{SO}(2)_\phi$-invariant linear response is block-diagonal in the $m$-basis, so an irreducible component of a bunch moment carrying azimuthal index $m$ couples only to the $m$-th harmonic of Eq.~\eqref{eq:multipole}. The two resulting selection rules are kinematic constraints from representation theory: they identify which $m$ must be populated for a given observable to respond, while the magnitude of the response coefficient $\chi_m$ is set by the nonlinear plasma dynamics and depends on the operating regime. A third statement (S3) below lies outside the $\mathrm{SO}(2)_\phi$ representation theory; it is a symplectic-form analogy included here for presentational convenience and labelled accordingly.

(S1) Only $m{=}1$ displaces the bunch centroid. The dipole moment of the bunch transforms under $\mathrm{SO}(2)_\phi$ as a vector, and only the $m{=}1$ harmonic of the wake contains such a vector. An axisymmetric ($m{=}0$) perturbation cannot move $\langle\boldsymbol{x}_\perp\rangle$ by any amount, no matter how strong. Centroid jitter, pointing noise, and hosing are therefore three names for the same channel. Hosing of the PWFA drive bunch~\citep{huang_hosing_2007,mehrling_mitigation_2017}, the analogous transverse modulation of the LWFA laser pulse~\citep{kaluza_observation_2010}, hosing of the witness~\citep{nechaeva_hosing_2024}, pulse-front-tilt jitter~\citep{seidel_polarization_2022}, and driver transverse misalignment all populate $\Psi_1^{c,s}$ and cannot be suppressed by axisymmetric means.

(S2) Only $m{=}2$ couples the two transverse planes. The $m{=}2$ harmonic of Eq.~\eqref{eq:multipole} has two components, $\Psi_2^c(r,\xi)\cos 2\phi \propto x^2 - y^2$ and $\Psi_2^s(r,\xi)\sin 2\phi \propto xy$; the two are related by a $\pi/4$ rotation of $(x,y)$ and describe the same elliptic perturbation in different axes. In the principal-axis frame the perturbed transverse focusing is anisotropic but separable, $\omega_x^2 x^2 + \omega_y^2 y^2$ with $\omega_x \neq \omega_y$; in any other frame an explicit $xy$ skew-focusing term appears and couples $(x,p_x)$ to $(y,p_y)$ in the equations of motion. The single-plane emittances $\varepsilon_n^{(x)}$ and $\varepsilon_n^{(y)}$ are no longer separately preserved, although the principal-axis emittances $\varepsilon_n^{(\pm)}$ remain symplectic invariants of the perturbed motion. Every cross-plane emittance-exchange mechanism in a plasma accelerator --- elliptic drivers~\citep{xu_low_2015,assmann_proton_2014}, beam-induced ion motion~\citep{diederichs_resonant_2025}, skew or misaligned quadrupoles in staging transfer lines~\citep{lindstrom_staging_2021} --- populates the $m{=}2$ tensor and lives in this channel.

Table~\ref{tab:m-inventory} collects the interpretation of each $m$ used throughout the remainder of the review.

\begin{table*}[!htbp]
  \centering
  \caption{Azimuthal-mode inventory for the blowout-wake perturbation expansion~\eqref{eq:multipole}. Each row corresponds to one multipole order $m$ and records the bunch moment it couples to, the experimentally observed manifestation, and the class of mitigation that breaks the feedback loop at that order. The $m{=}3$ row records the channel raised in Sec.~\ref{sec:conclusions}, whose response coefficient $\chi_3$ has no quantitative bound in the published literature.}
  \label{tab:m-inventory}
  \begin{ruledtabular}
  \begin{tabular}{c l p{0.35\linewidth} p{0.35\linewidth}}
    $m$ & Witness moment & Experimental manifestation & Mitigation class \\
    \colrule
    0 & charge, rms size $\sigma_x\sigma_y$ & accelerating + focusing (desired)~\citep{lu_nonlinear_2006} & --- \\
    1 & centroid $\langle\boldsymbol{x}_\perp\rangle$ & hosing, pointing jitter, PFT-induced slice offset & driver stabilisation; plasma-frequency detuning~\citep{moreira_mitigation_2023} \\
    2 & $\langle xy\rangle$, $\sigma_x/\sigma_y$ & transverse ellipticity, cross-plane emittance mixing & flat driver; ion-motion suppression~\citep{diederichs_resonant_2025} \\
    3 & $\langle x^3{-}3xy^2\rangle,\;\langle 3x^2y{-}y^3\rangle$ & hexapolar slice-dependent skewness, if populated & $m{=}3$-specific detuning, if needed \\
  \end{tabular}
  \end{ruledtabular}
\end{table*}

Separately, a symplectic-form analogy --- not a consequence of the $SO(2)$ representation theory above --- is worth recording because we use it in Sec.~\ref{subsec:beam-loading}.

(S3) Symplectic-form analogy between transverse matching and longitudinal beam loading. The co-moving frame carries a second canonical pair $(\xi, p_\xi)$ with longitudinal Hamiltonian $H_\xi(\xi, p_\xi)$ whose Hamilton equations share the functional form of those for $(x, p_x)$. Matching in the transverse plane and optimal beam loading in $\xi$ are both symplectic-preservation problems, and both fail through phase-space filamentation when mismatched. The analogy is formal: transverse matching is a single-particle linear optics problem driven by an external, bunch-independent ion column, whereas longitudinal beam loading is a collective, nonlinear problem sourced by the bunch's own current.

The multipole classification relies on the convergence of the expansion~\eqref{eq:multipole} around axisymmetry. This is a small-perturbation statement: it holds when all transverse bunch moments are small compared to the cavity scale $k_p^{-1}$. Flat collider beams with design aspect ratios of order $10^2$ sit outside this regime; the large transverse aspect ratio populates many $m$-harmonics at comparable amplitude and the linear-multipole expansion loses its discriminating power. The classification is therefore most useful for (i) all single-stage and staging contexts where the bunch is approximately round ($k_p \sigma_r \lesssim 0.3$), (ii) FEL-driver designs where round-beam optimisation dominates, and (iii) the injection and matching sections of a collider where the beam has not yet been flattened.

\subsection{Symplectic transport and the Lorentz-invariant \texorpdfstring{$\varepsilon_n$}{normalised emittance}}
\label{subsec:symplectic}

The transverse single-particle dynamics in the unperturbed ($m{=}0$) blowout cavity is generated by a time-dependent but quadratic Hamiltonian of Courant--Snyder form~\citep{courant_theory_1958,chao_physics_1993,reiser_theory_2008},
\begin{equation}
\label{eq:cs-hamiltonian}
H_{\perp}(\xi) \;=\; \frac{p_x^2 + p_y^2}{2\gamma} + \tfrac{1}{2}\gamma \,\omega_{\beta}^2(\xi)\,(x^2 + y^2)\,,
\end{equation}
where $\gamma$ is the single-particle Lorentz factor, $(p_x, p_y)$ the transverse canonical momenta, and $\omega_{\beta}(\xi) = k_p c /\sqrt{2\gamma}$ the betatron frequency in the parabolic pseudopotential $\Psi_0^c$ of Eq.~\eqref{eq:multipole}; the explicit $\xi$-dependence enters through the slice energy. Higher-$m$ contributions, including the $m{=}2$ cross-plane couplings of (S2), enter as perturbations to Eq.~\eqref{eq:cs-hamiltonian}. The transverse map is symplectic, and the Courant--Snyder action $J_x = \tfrac12(\gamma_{\rm CS} x^2 + 2\alpha_{\rm CS} x x' + \beta_{\rm CS} x'^2)$ --- with $\{\alpha_{\rm CS}, \beta_{\rm CS}, \gamma_{\rm CS}\}$ the Twiss parameters of the matched orbit and $x' \equiv dx/dz$ the divergence angle --- is an adiabatic invariant. The rms geometric emittance $\varepsilon = \sqrt{\langle x^2\rangle\langle x'^2\rangle - \langle xx'\rangle^2}$ and the normalised emittance $\varepsilon_n = \gamma\beta\,\varepsilon$ (with $\beta = v/c$ the relativistic velocity; not to be confused with the Twiss $\beta_{\rm CS}$) are both invariants of the linear motion. The normalised form is additionally Lorentz-invariant under boosts --- it depends only on the phase-space volume, which is preserved by any symplectic map including the longitudinal boost --- and is therefore the correct figure of merit for comparing beams at different energies~\citep{reiser_theory_2008}. The geometric (un-normalised) emittance damps adiabatically as $\varepsilon \propto 1/(\gamma\beta)$ during matched acceleration, while the normalised $\varepsilon_n$ just defined remains invariant.

A direct consequence of axisymmetry: in the cavity of Eq.~\eqref{eq:cs-hamiltonian}, the two single-plane actions $J_x$ and $J_y$ --- and hence $\varepsilon_n^{(x)}, \varepsilon_n^{(y)}$ --- are separately conserved at all orders in the linear-focusing approximation. Cross-plane emittance growth is therefore a diagnostic for broken axisymmetry; the physical mechanisms (skew quadrupoles upstream, elliptic drivers, beam-induced ion motion) are the $m{=}2$ channel classified by (S2).

When the Courant--Snyder + axisymmetric picture above fails, it fails through one of three identifiable channels: (i) chromaticity --- an energy-dependent $\omega_\beta$ that gives slices of different energy distinct betatron phase advance; (ii) nonlinear focusing --- an amplitude-dependent $\omega_\beta$ that violates the quadratic form of Eq.~\eqref{eq:cs-hamiltonian}; and (iii) $m{=}2$ cross-plane coupling. For laser-driven beams chromaticity is typically the most aggressive of the three. The matched beta function inside the plasma channel, $\beta^* \sim \sqrt{2\gamma}/k_p$, sits at the sub-micron level for sub-GeV beams in a $10^{18}~\text{cm}^{-3}$ plasma; this small $\beta^*$ converts even sub-per-cent energy spreads into substantial betatron-frequency dispersion across the bunch over sub-millimetre propagation distances.

\begin{figure}[!htbp]
  \centering
  \includegraphics[width=0.9\linewidth]{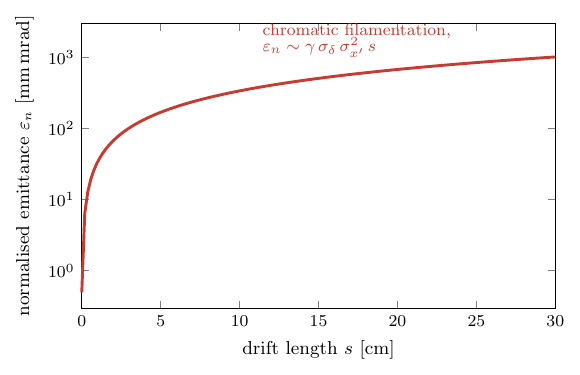}
  \caption{Normalised emittance $\varepsilon_n$ along an unmatched vacuum drift downstream of the plasma exit, adapted from Ref.~\citep{migliorati_intrinsic_2013}: the analytic chromatic-filamentation model (their Eq.~(5)) at multi-hundred-MeV laser-plasma-born bunch parameters. The growth is linear in the drift length $s$; the apparent curvature is a logarithmic-ordinate artefact. The plasma-exit value $\varepsilon_n \approx 0.5$~mm$\cdot$mrad would be preserved by matched transport; the figure plots departure from that baseline. See the surrounding text for the physical interpretation and the implications for plasma-exit matching.}
  \label{fig:migliorati}
\end{figure}

Figure~\ref{fig:migliorati} makes this quantitative for an LWFA-born bunch transported through an unmatched vacuum drift downstream of the plasma exit. The plotted curve is the analytic chromatic-filamentation model of~\citet{migliorati_intrinsic_2013} (their Eq.~(5)), $\varepsilon_n(s) \sim \gamma\,\sigma_\delta\,\sigma_{x'}^2\,s$, evaluated at multi-hundred-MeV bunch parameters with $\sim 6\%$ energy spread; the plasma-exit value $\varepsilon_n \approx 0.5~\text{mm}\cdot\text{mrad}$ would be preserved by matched transport. The normalised emittance grows by roughly three orders of magnitude over tens of centimetres. The growth has a single-particle origin: slices of different energy accumulate different betatron phase advances during the unmatched drift, and the resulting shear of the transverse distribution inflates the rms emittance even though the single-particle motion remains symplectic. This is the quantitative motivation for immediate-capture matching at the plasma exit (Sec.~\ref{subsec:matching}).

\section{LWFA beam quality}
\label{sec:lwfa}

We apply the selection rules of Sec.~\ref{subsec:selection-rules} to the laser-driven case. Section~\ref{subsec:injection} treats injection schemes as instances of controlled $T_\xi^{\rm slow}$ breaking; Section~\ref{subsec:m1-lwfa} collects LWFA mechanisms that excite the $m{=}1$ channel; Section~\ref{subsec:m2-lwfa} collects those exciting $m{=}2$; Section~\ref{subsec:spin} covers spin polarisation, whose dynamics decouple from the multipole sector through Thomas--BMT precession.

\subsection{Injection as controlled breaking of longitudinal translation}
\label{subsec:injection}

Injection schemes all work by breaking $T_\xi^{\rm slow}$ at a well-chosen place along the accelerator. Density-downramp injection~\citep{geddes_plasma_2008} lowers the wake phase velocity by changing the plasma wavelength over a short distance, allowing trapping into the otherwise closed separatrix. Ionisation injection~\citep{pak_injection_2010} breaks the envelope-level time translation: K-shell electrons of a dopant species are released only at the peak of the driver field, at a $\xi$ determined by the ionisation threshold. Colliding-pulse schemes~\citep{faure_controlled_2006} break both $T_\xi^{\rm slow}$ and $\mathrm{SO}(2)_{\phi}$ through the transient beat wave. Self-injection~\citep{kalmykov_electron_2009} breaks $T_\xi^{\rm slow}$ spontaneously: the driver's own evolution makes the bubble shape unsteady on the bunch transit timescale, opening the trapping separatrix without any externally imposed perturbation.

The symmetries broken at injection time organise the phase-space signature that each scheme would imprint on the witness. A downramp pins longitudinal phase but leaves $\mathrm{SO}(2)_{\phi}$ intact, so an injected bunch inherits the axisymmetry of the cavity at the injection point. Colliding-pulse breaks $\mathrm{SO}(2)_{\phi}$ transiently, so the injected bunch should carry a transverse anisotropy correlated with the polarisation of the colliding pulse. Ionisation injection leaves both $T_\xi^{\rm slow}$ and $\mathrm{SO}(2)_{\phi}$ largely intact after the ionisation event, so injection contributes only a narrow trap-phase spread and no injection-time ellipticity. Direct experimental tests of these inheritance patterns are sparse: routine LWFA diagnostics report charge, energy, and divergence rather than the separate single-plane emittances or polarisation-correlated anisotropies that would probe them, and downstream symmetry-breaking sources (driver evolution, beam loading, transport optics) can mask or overwrite the injection-time imprint. Three recent developments sharpen the empirical picture: controlled downramps in meter-scale plasma~\citep{hue_control_2025}, guided-mode ionisation injection at multi-GeV~\citep{shrock_guided_2023}, and resonant multi-pulse ionisation~\citep{tomassini_resonant_2017}. Table~\ref{tab:injection} collects the LWFA and PWFA injection mechanisms by the symmetries each breaks.

\begin{table*}[!htbp]
  \centering
  \caption{Injection schemes in LWFA and PWFA, classified by which symmetries of the unperturbed wake~\eqref{eq:symmetry-group} they break at the injection point. The signature in phase space follows from the conservation laws that remain intact after injection: axisymmetry preservation implies $\langle xy\rangle=0$ at release, longitudinal-translation breaking pins the trap phase, and combined breaking pins both.}
  \label{tab:injection}
  \begin{ruledtabular}
  \begin{tabular}{l l p{0.40\linewidth} l}
    Scheme & Regime & Broken symmetries & Key refs \\
    \colrule
    Density downramp & LWFA, PWFA & $T_\xi^{\rm slow}$ & \citep{geddes_plasma_2008,hue_control_2025,ossa_wakefield_2015} \\
    Ionisation injection & LWFA & $T_\xi^{\rm slow}$ (envelope) & \citep{pak_injection_2010,shrock_guided_2023,tomassini_resonant_2017} \\
    Wakefield-induced ionisation & PWFA & $T_\xi^{\rm slow}$ (drive peak) & \citep{ossa_wakefield_2015,li_generation_2015} \\
    Colliding-pulse & LWFA & $T_\xi^{\rm slow}+\mathrm{SO}(2)_{\phi}$ (transient) & \citep{faure_controlled_2006} \\
    Self-injection & LWFA & $T_\xi^{\rm slow}$ (spontaneous, via driver evolution) & \citep{kalmykov_electron_2009} \\
    Plasma photocathode & PWFA & $\mathrm{SO}(2)_{\phi}$ (optical trigger) & \citep{li_high_2018} \\
  \end{tabular}
  \end{ruledtabular}
\end{table*}

\subsection{The \texorpdfstring{$m{=}1$}{m=1} channel in LWFA}
\label{subsec:m1-lwfa}

Any process that displaces the bunch centroid in the transverse plane populates the $m{=}1$ amplitude $\Psi_1^{c,s}$ of Eq.~\eqref{eq:multipole}. By (S1), no axisymmetric cause can do this; every $m{=}1$ excitation in LWFA is traceable to a specific broken-axisymmetry input. Driver transverse misalignment or pointing jitter is the obvious candidate, but several less-obvious inputs contribute: pulse-front tilt writes a slice-dependent $m{=}1$ offset along the bunch~\citep{seidel_polarization_2022}; laser polarisation couples to residual bubble anisotropy to produce a polarisation-dependent centroid~\citep{seidel_polarization_2022}; and the laser-hose instability itself is a coherent $m{=}1$ mode of the driver--wake system, observed experimentally as long-wavelength centroid modulation of the laser pulse~\citep{kaluza_observation_2010}, with the bunch-driver analogue treated separately in Sec.~\ref{subsec:pwfa-hosing}. Mitigation strategies act either by suppressing the input (active driver stabilisation) or by breaking the coherent feedback loop (plasma-frequency detuning~\citep{moreira_mitigation_2023}). These are distinct physical interventions with the same symmetry effect: they prevent the $m{=}1$ coefficient from accumulating phase.

\subsection{The \texorpdfstring{$m{=}2$}{m=2} channel in LWFA}
\label{subsec:m2-lwfa}

Quadrupolar perturbations populate $\Psi_2^{c,s}$. In LWFA the dominant sources are (i) elliptic laser envelopes from astigmatism or non-ideal focal-spot symmetry, which seed the same $m{=}2$ wake structure analysed for elliptic drive bunches in~\citet{manwani_analysis_2024}; (ii) linear laser polarisation, whose contribution to the transverse ponderomotive force is small after cycle-averaging but cumulative over many betatron periods~\citep{seidel_polarization_2022,kaluza_observation_2010}; and (iii) asymmetric plasma channels. By (S2), any cross-plane emittance mixing seen in LWFA is in this channel, independent of which of the three inputs is responsible; direct laser acceleration (DLA)~\citep{shaw_role_2015,zhang_synergistic_2015} becomes observable when the $m{=}2$ component of the bubble transverse field resonates with the driver phase velocity and accelerates a subset of witness electrons longitudinally with a specific angular signature. The practical consequence is that the $m{=}2$ channel has both a coupling effect (emittance mixing) and a dispersion effect (DLA contamination of the energy spectrum) -- the two are co-measured.

\subsection{Spin polarisation and Thomas--BMT precession}
\label{subsec:spin}

The spin of a single electron is an internal SU(2) degree of freedom; its evolution in the wake is governed by the Thomas--BMT equation~\citep{bargmann_precession_1959} with the local magnetic field and the centripetal acceleration. Writing $a\equiv(g{-}2)/2 \approx 1.16\times 10^{-3}$ for the electron anomalous magnetic moment and $\omega_\beta$ for the betatron angular frequency in the wake, the anomalous spin-precession rate relative to the momentum vector is $\gamma a\,\omega_\beta$, which at GeV energies ($\gamma\sim 10^3$) gives $\gamma a \approx 1.16$ and is therefore of order $\omega_\beta$ itself --- the spin and momentum vectors detune by $\mathcal{O}(1)$ radian per betatron oscillation. The consequence is that ``spin preserved by the wake'' is a statement that depends on specifying which axis and over what length.

Within the $m$-channel framework, the spin dynamics decomposes along the same multipoles as the beam moments. The axisymmetric ($m{=}0$) magnetic field drives pure Larmor precession about the drive axis~\citep{bargmann_precession_1959}: a spin polarisation aligned with the axis is preserved exactly, and a transverse polarisation precesses coherently at the Larmor rate with no depolarisation of the ensemble. The $m{=}1$ harmonic of the wake magnetic field is driven by the same centroid displacement that underlies the hose instability~\citep{kaluza_observation_2010,huang_hosing_2007}, and produces a correlated spin tilt along the bunch. The $m{=}2$ harmonic produces anisotropic precession rates that decohere the transverse spin across the ensemble on the wake-induced emittance-mixing timescale~\citep{diederichs_resonant_2025}. Depolarisation is therefore controlled by the same $m$-spectrum as emittance growth: symmetry-preserving wakes preserve polarisation automatically, and the mitigations that suppress the $m{=}1$ and $m{=}2$ channels also protect the spin state.

This framework maps directly onto experimental polarised-injection studies. Pre-polarised plasma sources from HCl or CH$_3$I photo-dissociation supply polarised electrons at injection, and LWFA preserves polarisation along the driver axis with demonstrated efficiencies above 80\%~\citep{wen_polarized_2019,wen_generation_2020}. The beam-driven analogue uses plasma-photocathode injection in the drive-bunch field and reaches multi-GeV~\citep{nie_spinpolarized_2021,nie_highly_2022}. Vortex-laser drivers shape the output spin state through orbital-spin coupling of the OAM mode~\citep{wu_polarized_2019}, and in the QED regime radiation reaction itself produces polarisation of initially unpolarised beams through the Sokolov--Ternov mechanism~\citep{delsorbo_polarization_2019}. Polarisation control in the bubble regime~\citep{fan_control_2022} and colliding-pulse polarised injection~\citep{sun_colliding_2023} extend the same idea: the injection step is engineered as a deliberate symmetry-breaking operation that selects a specific spin state of the trapped population.

\section{PWFA beam quality}
\label{sec:pwfa}

Section~\ref{sec:pwfa} transposes the framework to the beam-driven case. The symmetry group~\eqref{eq:symmetry-group} and the multipole expansion~\eqref{eq:multipole} are unchanged: the driver is a charged bunch rather than a laser pulse, but the wake-cavity geometry and its symmetries are identical. What changes is which inputs populate which azimuthal multipole order $m$ --- the integer labelling the irreducible representations of $\mathrm{SO}(2)_\phi$ introduced in Sec.~\ref{sec:framework}, with $m{=}1$ governing centroid motion, $m{=}2$ cross-plane coupling, and $m{=}0$ the unperturbed axisymmetric cavity.

\subsection{Witness injection}
\label{subsec:pwfa-injection}

Wakefield-induced ionisation~\citep{ossa_wakefield_2015,li_generation_2015} releases witness electrons from a dopant gas at the peak field of the drive bunch and plays the same role as laser-driven ionisation injection: breaking of $T_\xi^{\rm slow}$ at the ionisation threshold. The transverse-momentum kick at birth is different --- in a laser pulse it is set by the normalised laser vector potential $a_0 \equiv eA_0/(m_e c)$ at the ionisation time, with $A_0$ the peak vector-potential amplitude and $e$, $m_e$ the electron charge and rest mass; in a bunch-driver wake it is set by the transverse bunch field profile --- so the correspondence is not complete. All-optical density-downramp injection in beam-driven wakes~\citep{ossa_wakefield_2015} completes the LWFA$\leftrightarrow$PWFA analogy at the longitudinal level, while the plasma-photocathode scheme~\citep{li_high_2018} is the cleanest example of a transverse-symmetry-breaking injection: a weak optical pulse breaks $\mathrm{SO}(2)_{\phi}$ at the injection point and produces sub-$\mu$m normalised emittance witness beams that would be forbidden by the (S2) selection rule in a strictly axisymmetric setup.

\subsection{The \texorpdfstring{$m{=}1$}{m=1} channel in PWFA}
\label{subsec:pwfa-hosing}

Bunch-driver hosing~\citep{huang_hosing_2007,mehrling_mitigation_2017} --- the hose instability (HI) of the drive bunch --- is the $m{=}1$ analogue of LWFA laser-driver hosing with the drive-bunch centroid replacing the laser centroid. The underlying selection rule (S1) is the same and so are the mitigations: plasma-frequency detuning~\citep{moreira_mitigation_2023} destroys the phase-locking that lets the coherent $m{=}1$ mode grow. Long proton drivers couple to the same $m{=}1$ channel through the self-modulation instability (SMI)~\citep{assmann_proton_2014}, which saturates once the bunch is modulated on the plasma scale and transfers into the $m{=}0$ drive channel. Short-range wakefields in the blowout~\citep{stupakov_short_2018} place a hard lower bound on witness--drive separation consistent with $m{=}1$ control.

\begin{figure*}[!htbp]
  \centering
  \includegraphics[width=0.55\linewidth]{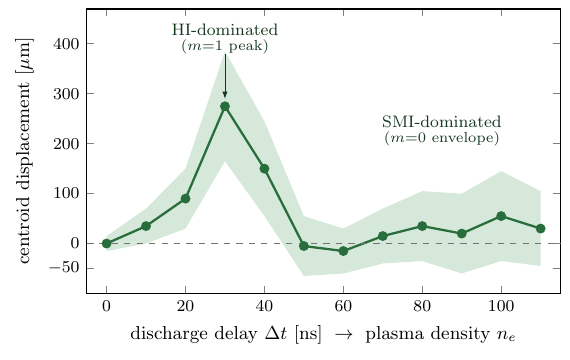}
  \caption{Beam centroid displacement of a long drive bunch as a function of the discharge delay $\Delta t$, reproduced from Fig.~4(a) of Ref.~\citep{del_dotto_experimental_2022}; $\Delta t$ sets the plasma density $n_e$ at the time of beam arrival. Points are read off the published figure, the shaded band is the reported $1\sigma$ shot-to-shot spread, and the dashed line is the plasma-off reference. The annotated regimes are interpreted in the main text; the companion tilt-angle data and the supporting PIC simulations are in the original reference.}
  \label{fig:hi-smi-transition}
\end{figure*}

Under (S1), HI and SMI are two regimes of the same long-bunch wake rather than independent instabilities: which channel dominates is selected by where the plasma scale $k_p^{-1}$ sits relative to the bunch length $\sigma_z$. At $k_p \sigma_z \lesssim 1$ the centroid-driven $m{=}1$ mode dominates; at $k_p \sigma_z \gtrsim 1$ the envelope-driven $m{=}0$ mode takes over. The density scan reported at SPARC\_LAB~\citep{del_dotto_experimental_2022}, reproduced in Fig.~\ref{fig:hi-smi-transition}, demonstrates this $m$-selection directly: the $m{=}1$ centroid displacement peaks near $\Delta t = 30$~ns and is progressively suppressed as larger delays raise the plasma density into the $m{=}0$-dominated regime visible at $\Delta t \gtrsim 50$~ns in the figure. The density axis thereby acts as the $m$-selection control parameter of Sec.~\ref{sec:framework} within a single long-bunch wake.

\subsection{Beam loading as longitudinal matching}
\label{subsec:beam-loading}

Optimal beam loading flattens the longitudinal accelerating field $E_z(\xi)$ across the witness length, fixing the slice energy gain to a common value and arresting growth of the correlated energy spread. Following the symplectic-form analogy (S3), the matched longitudinal condition is the $\xi$-plane counterpart of the transverse matching condition: it minimises correlated-energy-spread growth in the longitudinal plane the way matched transverse injection minimises chromaticity-driven emittance growth in the transverse plane~\citep{ariniello_chromatic_2022}. The two problems differ in causality --- transverse matching fights a bunch-independent ion-column focusing, longitudinal loading shapes the bunch's own contribution to the accelerating field --- but both fail through phase-space filamentation when mismatched, and both are governed by the same symplectic-invariance constraints. Measured sub-per-cent energy spreads at FACET-II~\citep{pompili_energy_2021,lindstrom_emittance_2024} demonstrate that the matched longitudinal condition is reachable in practice. Shaped drive currents enhance the transformer ratio~\citep{roussel_single_2020} only when they preserve the matched longitudinal condition; transformer ratios obtained without matched loading produce large correlated energy spread and do not translate into useful brightness for FEL or collider applications.

\subsection{Positron witnesses and the parity asymmetry}
\label{subsec:positron}

The blowout cavity generated by an electron driver focuses electrons but defocuses positrons~\citep{gjcao_positron_2024}. The asymmetry follows from Gauss's law applied to the uniform positive ion background of the blowout: inside the fully evacuated cavity the radial field obeys $E_r \propto r$, so the transverse force $-eE_r$ confines negative charges and expels positive ones, with the focusing strength set by the ion density $n_0$ alone and independent of the driver geometry. The discrete parity $P_{z\leftrightarrow -z}$ of Eq.~\eqref{eq:symmetry-group} exchanges witness charge signs and the effective focusing channel transforms with opposite sign under this operation. The corresponding selection rule is restrictive: any perturbation of the wake that preserves both the uniform ion density and the fully evacuated ion column leaves the $E_r \propto r$ focusing intact, so no such perturbation can yield charge-sign-symmetric focusing.

Each demonstrated positron-focusing scheme breaks one of these two features (Table~\ref{tab:positron}). Hollow plasma channels~\citep{li_high_2018} replace the on-axis plasma with a vacuum core surrounded by an annular plasma; axisymmetry is preserved, but the linear ion focusing of the uniform blowout is replaced by the response of the annular plasma boundary, which can be made positron-focusing for a suitably shaped drive bunch. Tailored-density-profile schemes~\citep{diederichs_emittance_2023} retain axisymmetry but break the uniform-column assumption: a residual electron filament survives on the axis and provides partial positron focusing, with finite plasma temperature linearising the focusing field within a narrow on-axis region. Flat or elliptic drivers~\citep{gjcao_positron_2024} deliberately populate the $m{=}2$ harmonic of the wake, breaking $\mathrm{SO}(2)_\phi$ of the drive while preserving the uniform ion density; the resulting quadrupolar focusing acts on positrons in one transverse plane and on electrons in the orthogonal one. The three schemes correspond to relaxing, respectively, full evacuation of the ion column, uniformity of the ion density, and axisymmetry of the drive.

Multipole superpositions within the uniform-density blowout do not constitute a separate scheme. Each $m \ge 1$ multipole field of Eq.~\eqref{eq:multipole} derives from a harmonic potential $r^m \cos m\phi$ that contributes no $m{=}0$ component to the focusing, so any linear combination of $m \ge 1$ perturbations leaves the Gauss's-law-determined $m{=}0$ focusing set by the uniform ion density unchanged. The three schemes of Table~\ref{tab:positron} therefore exhaust the perturbative routes within the framework of Eq.~\eqref{eq:symmetry-group} that preserve the electrostatic plasma response; schemes that supplement this response with external static fields lie outside the framework and are not addressed here.

\begin{table*}[!htbp]
  \centering
  \caption{Positron-focusing schemes in PWFA, classified by the feature of the uniform-density electron-driven blowout each scheme relaxes. Hollow channels preserve axisymmetry but abandon the full ion column; warm-filament wakes preserve axisymmetry but tailor the density profile; asymmetric drivers preserve uniformity but break $\mathrm{SO}(2)_\phi$ of the drive geometry. Residual issues are the principal obstructions identified in the published demonstrations.}
  \label{tab:positron}
  \begin{ruledtabular}
  \begin{tabular}{l p{0.27\linewidth} p{0.32\linewidth} l}
    Scheme & Feature of uniform blowout relaxed & Residual issue & Key refs \\
    \colrule
    Hollow plasma channel & full ion column (vacuum core; $\mathrm{SO}(2)_\phi$ preserved) & transverse beam-breakup from drive misalignment; absence of on-axis focusing in the cold-channel limit & \citep{li_high_2018,lindstrom_measurement_2018,silva_stable_2021} \\
    Warm-filament wake & uniform ion density (tailored profile; $\mathrm{SO}(2)_\phi$ preserved) & non-linear positron focusing outside the narrow on-axis region linearised by finite plasma temperature & \citep{diederichs_emittance_2023} \\
    Asymmetric drive (flat / elliptic) & $\mathrm{SO}(2)_\phi$ of the drive geometry (deliberately populates $m{=}2$) & matching a round witness into the elliptic focusing & \citep{gjcao_positron_2024,manwani_analysis_2024} \\
  \end{tabular}
  \end{ruledtabular}
\end{table*}

\section{Emittance preservation as a symplectic problem}
\label{sec:emittance}

We collect the emittance-preservation literature here and apply the organising rules of Sec.~\ref{sec:framework} to identify the hose instability and the resonant cross-plane mixing reported by~\citet{diederichs_resonant_2025} as two members of a multipole-indexed classification. The shared structure lies in the multipole geometry of the feedback loop at each $m$, while the physics of the response coefficient $\chi_m$ that closes the loop differs across $m$, as set out below.

\subsection{Chromaticity, nonlinearity, and coupling}
\label{subsec:symplectic-transport}

Linear symplectic transport in an axisymmetric channel preserves $\varepsilon_n^{(x)}$ and $\varepsilon_n^{(y)}$ separately, as established in Sec.~\ref{subsec:symplectic}. Growth must therefore come from one of three identifiable deviations: (i) chromatic spread of the betatron frequency across the bunch energy distribution, which leaves axisymmetry intact but introduces slice-dependent phase advance; (ii) amplitude-dependent nonlinear focusing, which leaves axisymmetry intact but drives rms emittance growth through phase-space filamentation~\citep{migliorati_intrinsic_2013}; (iii) $m{=}2$ cross-plane coupling. In staged plasma accelerators, chromatic growth is typically dominant: the strong plasma focusing amplifies even sub-per-cent energy spreads into measurable betatron-frequency dispersion across the bunch~\citep{lindstrom_emittance_2022,ferrario_injection_2020,ariniello_chromatic_2022}. The same mechanism is already operative at the plasma--vacuum interface, where the abrupt loss of focusing combined with the sub-micron $\beta^*$ produces rapid, chromaticity-driven growth of the rms transverse emittance over the first millimetres of free drift~\citep{migliorati_intrinsic_2013} --- the single-particle phase-mixing effect quantified in Fig.~\ref{fig:migliorati}, and the quantitative motivation for immediate-capture matching (Sec.~\ref{subsec:matching}). The cross-plane coupling contribution in mechanism~(iii) is treated in Sec.~\ref{subsec:mixing}.

\subsection{Matching and active plasma lenses}
\label{subsec:matching}

An active plasma lens provides azimuthally symmetric, linear focusing and, in the aberration-free limit, preserves emittance through extraction and stage matching~\citep{rockemann_direct_2018,pompili_plasma_2020}. Exact phase-space matching protocols between conventional and plasma sections~\citep{xu_exact_2016,olsen_emittance_2018} keep mismatch-induced filamentation under control. Under the (S3) duality, the matching condition is the transverse counterpart of optimal beam loading in Sec.~\ref{subsec:beam-loading}: both fix the bunch to a specific operating point in the relevant phase plane and both fail through phase-space filamentation under departures from that point.

\subsection{Hosing and the \texorpdfstring{$m{=}2$}{m=2} mixing resonance}
\label{subsec:mixing}

\begin{figure*}[!htbp]
  \centering
  \includegraphics[width=0.55\linewidth]{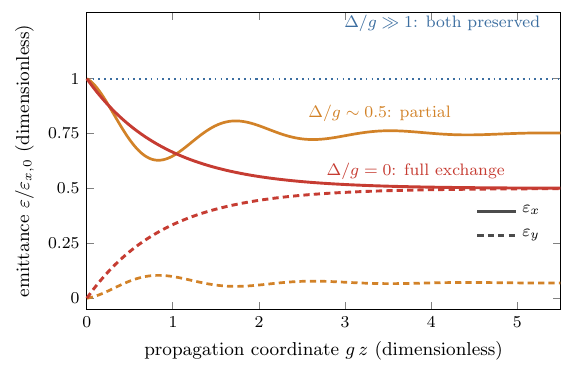}
  \caption{Adapted from Ref.~\citep{diederichs_resonant_2025}. Horizontal witness emittance $\varepsilon_x$ (solid) and vertical witness emittance $\varepsilon_y$ (dashed), normalised to the initial horizontal emittance $\varepsilon_{x,0}$ (dimensionless), along the dimensionless propagation coordinate $g\,z$, with $g$ the coupling strength between the two transverse planes. The control parameter $\Delta/g$ is the ratio of the transverse betatron-frequency mismatch $\Delta \equiv \omega_{\beta,x}-\omega_{\beta,y}$ to the coupling strength $g$, evaluated here at $\Delta/g = 0$ (red), $\Delta/g \sim 0.5$ (orange), and $\Delta/g \gg 1$ (blue). The three regimes are interpreted in the main text below. Curves are drawn from an ensemble-averaged Rabi-exchange model; see Ref.~\citep{diederichs_resonant_2025} for the original quasi-static 3D PIC demonstration.}
  \label{fig:diederichs-mixing}
\end{figure*}

The multipole structure of the wake organises the dominant emittance-growth instabilities of the blowout regime by their azimuthal index. A witness whose $m$-th transverse moment is nonzero excites the $m$-th harmonic of the wake, which acts back on the same moment and closes a feedback loop at that order. At $m{=}1$ this loop is the classical drive-beam and witness hosing~\citep{whittum_electron_1991,huang_hosing_2007,schroeder_hose_2012,mehrling_mitigation_2017,nechaeva_hosing_2024}: a centroid displacement drives a dipole wake that further displaces the centroid. At $m{=}2$, the same geometric picture gives the resonant cross-plane mixing reported in~\citep{diederichs_resonant_2025}: quadrupolar ellipticity of the witness drives an $m{=}2$ wake response through nonlinear ion motion or beam-induced ionisation, which then exchanges emittance between the two transverse planes. Figure~\ref{fig:diederichs-mixing} reproduces the quantitative demonstration of the $m{=}2$ loop and its suppression by detuning the two transverse betatron frequencies.

At $\Delta/g = 0$ (red curves) the two transverse betatron frequencies are matched, the coherence condition for the $m{=}2$ loop is satisfied, and $\varepsilon_x$ and $\varepsilon_y$ saturate at $\tfrac12 \varepsilon_{x,0}$ through a full Rabi-like ensemble exchange. At $\Delta/g \sim 0.5$ (orange) only a fraction of the witness particles are resonant; the partial exchange saturates at a higher residual $\varepsilon_x$ and a small but nonzero $\varepsilon_y$, both carrying an oscillation from the incomplete averaging over the partially resonant sub-population. At $\Delta/g \gg 1$ (blue, dotted reference) the loop is broken and both emittances remain at their initial values. The dimensionless detuning $\Delta/g$ is therefore the control parameter of the $m{=}2$ loop, playing at $m{=}2$ the role that the Balakin--Novokhatsky--Smirnov (BNS) damping recipe~\citep{balakin_bns_1983} plays at $m{=}1$ in suppressing hosing.

The two loops are at different stages of experimental development. The $m{=}1$ branch is empirically established: long-bunch electron-hose theory was developed in~\citep{whittum_electron_1991,schroeder_hose_2012}, the hose-to-self-modulation transition has been directly measured~\citep{del_dotto_experimental_2022} (Fig.~\ref{fig:hi-smi-transition}), the saturation of self-modulation as a hose-suppression mechanism has been predicted and observed~\citep{vieira_hosing_2014,turner_experimental_2019}, and witness-beam hosing is routinely characterised in PWFA simulations against this backdrop. To our knowledge the $m{=}2$ mixing branch has not yet been observed experimentally, and is so far supported by the quasi-static 3D PIC simulations of~\citep{diederichs_resonant_2025}. The asymmetry mirrors the $m{=}3$ gap discussed in Sec.~\ref{sec:conclusions}. Direct measurement of the $m{=}2$ coupling is a near-term experimental target.

The $m{=}1$ and $m{=}2$ loops share the same multipole geometry, but the response coefficient $\chi_m$ that closes the loop has a different physical origin at each $m$. At $m{=}1$, $\chi_1$ is a coherent dipole wake that propagates dynamically along the bunch and admits exponential growth; the characteristic instability length is set by the bunch--wake coupling~\citep{huang_hosing_2007,whittum_electron_1991,schroeder_hose_2012} and is of order centimetres at typical PWFA densities. At $m{=}2$, $\chi_2$ vanishes in the ideal blowout --- an axisymmetric ion column supports no $m{=}2$ response --- and must be populated by nonlinear physics beyond leading order (relativistic ion motion, beam-induced ionisation). In that picture the $m{=}2$ coupling acts as a transverse action-exchange resonance of single-particle dynamics at the betatron-frequency coincidence $k_{\beta,x}\approx k_{\beta,y}$~\citep{diederichs_resonant_2025} rather than as a coherent wave along the bunch. The multipole organisation pairs the two phenomena geometrically; the dynamical content of $\chi_m$ is then supplied separately by the plasma physics at each $m$.

Known suppression strategies at the two orders act by breaking a coherence condition of the loop, but the relevant coherence conditions are physically distinct. BNS damping~\citep{balakin_bns_1983} and plasma-frequency detuning~\citep{moreira_mitigation_2023,schroeder_hose_2012} suppress hosing by destroying the phase-locking of a fluid-like wake mode that propagates convectively along the bunch~\citep{huang_hosing_2007}. The flat-drive-beam scheme proposed in~\citep{diederichs_resonant_2025} suppresses the $m{=}2$ mixing by lifting a single-particle betatron-frequency degeneracy in the transverse plane through asymmetric ion motion. The multipole organisation identifies which order $m$ a suppression scheme must target, even though the specific control knob --- longitudinal-phase detuning at $m{=}1$, transverse-frequency detuning at $m{=}2$ --- changes with $m$.

A scope caveat closes the discussion. The multipole-feedback picture above is a small-perturbation statement in the witness moments, applied within the blowout regime: it requires all transverse witness moments to be small compared to the cavity scale $k_p^{-1}$, so that the wake response in each $m$ is well approximated by an independent multipole. Once any single channel saturates --- a fully developed $m{=}1$ hose deforms the cavity enough to seed $m{=}2,3,\ldots$ harmonics through cross-talk, and the $m{=}2$ mixing is itself driven by the nonlinear part of the plasma response --- the channels couple directly and the organisation must be supplemented with full nonlinear PIC accounting. The regime of interest for FEL-driver and collider-staging budgets, where the per-stage emittance growth tolerated by the design is a small fraction of the unperturbed emittance, sits within this small-perturbation regime.

\subsection{Staging and the per-stage emittance budget}
\label{subsec:staging}

Collider-scale designs chain many plasma stages, and the total emittance budget of the machine compounds the per-stage emittance growth across every matched transfer~\citep{lindstrom_emittance_2022,schroeder_beam_2022}. Recent 10~TeV pCM wakefield collider designs~\citep{gessner_design_2025} and the European Strategy accelerator R\&D roadmap~\citep{adolphsen_european_2022} identify per-stage $m{=}1$ and $m{=}2$ suppression as the dominant engineering constraint. Chromaticity, which preserves axisymmetry, can be handled stage by stage with the matching machinery of Sec.~\ref{subsec:matching}; cross-plane coupling at $m{=}2$, by contrast, compounds multiplicatively across stages.

\section{Conclusions and outlook}
\label{sec:conclusions}

We have organised the beam-quality literature for laser- and beam-driven plasma wakefield accelerators around the azimuthal-mode index $m$ assigned by the symmetry group of the idealised blowout wake. The three rules of Sec.~\ref{sec:framework} --- $m$-indexing of every beam-quality observable (S1), restriction of cross-plane coupling to $m{=}2$ (S2), and the symplectic correspondence between transverse matching and longitudinal beam loading (S3) --- provide a common basis on which the hose instability, pulse-front-tilt jitter, spot-asymmetry emittance growth, polarisation-dependent centroid motion, and the resonant cross-plane mixing reported in~\citep{diederichs_resonant_2025} appear as projections onto the two lowest non-trivial harmonics. The hose loop at $m{=}1$ and the mixing resonance at $m{=}2$ share the same multipole geometry and admit suppression schemes of the same form (frequency-domain detuning of the relevant channel), even though they are driven by physically distinct response coefficients $\chi_1$ and $\chi_2$. The positron-witness mitigation literature acquires a parallel structure: each established scheme corresponds to relaxing one specific symmetry of the uniform-density blowout, and the catalogue of available relaxations is finite.

Two directions for further work follow naturally. The first concerns the $m{=}3$ channel, which is permitted by $\mathrm{SO}(2)_\phi$ representation theory but whose response coefficient has not, to our knowledge, been characterised quantitatively in the published literature. As at $m{=}2$, the ideal axisymmetric blowout supports no $m{=}3$ response, and any nonzero contribution arises from nonlinear plasma physics. When the nonlinear source is analytic in the transverse displacement, as for relativistic ion motion, dimensional analysis gives $\chi_3/\chi_2 \sim k_p\sigma_r$; beam-induced ionisation, whose dependence on the field amplitude is threshold-like and therefore non-analytic, falls outside this estimate and is consequently the most informative regime in which to seek a measurable $m{=}3$ response. Kinematic suppression by the smallness of the witness radius relative to $k_p^{-1}$ makes the channel most relevant for narrow, approximately round witnesses sustained over long propagation distances --- precisely the regime of the injection and inter-stage sections of multi-stage collider designs~\citep{gessner_design_2025,adolphsen_european_2022}. Slice-resolved phase-space diagnostics already operational at current plasma-accelerator facilities report transverse moments up to second order; reconstruction of the third moments listed in Table~\ref{tab:m-inventory} is within reach as an extension of the existing analysis software. Full 3D particle-in-cell simulations resolve all azimuthal harmonics by construction, so a quantitative bound on the $m{=}3$ response from simulation is equally accessible. Establishing such a bound is a natural target in view of the per-stage emittance budgets that drive collider-scale designs.

The second direction concerns the interface with Bayesian optimisation, which is increasingly applied to the design and tuning of plasma accelerators in both simulation and experiment~\citep{shalloo_automation_2020,roussel_bayesian_2024,irshad_pareto_2024}. A generic Gaussian-process surrogate uses an isotropic covariance over the input parameters and recovers the symmetries of the underlying physics only from data. Encoding the $\mathrm{SO}(2)_\phi$ symmetry of the wake directly as an equivariance condition on the kernel would in principle reduce the effective search dimension; the selection rules of Sec.~\ref{subsec:selection-rules} indicate which input parameters preserve axisymmetry, which break it into the $m{=}1$ centroid channel, and which break it into the $m{=}2$ cross-plane channel, and could supply the corresponding structural priors. A systematic implementation of such priors in plasma-accelerator design pipelines, together with a treatment of the $\mathrm{P}_{z\leftrightarrow -z}$-broken positron-witness case (Sec.~\ref{subsec:positron}), is a natural avenue for future work.

\begin{acknowledgments}
A.C.B.\ acknowledges support from contracts PN23210105 and ELI-RO/RDI/2024\_040 funded by the Romanian Ministry of Education and Scientific Research, and the Extreme Light Infrastructure Nuclear Physics Phase II project, co-financed by the Romanian Government and the European Union through the European Regional Development Fund and the Competitiveness Operational Programme (No.\ 1/07.07.2016, COP, ID 1334). Large language model tools were used solely for language polishing; no content or data were generated by AI.
\end{acknowledgments}

\paragraph*{Author Contributions.}
Conceptualization, A.C.B.\ and A.D.D.; formal analysis, A.C.B.\ and A.D.D.; investigation, A.C.B.\ and A.D.D.; writing---original draft preparation, A.C.B.; writing---review and editing, A.C.B.\ and A.D.D.; visualization, A.C.B. All authors have read and agreed to the published version of the manuscript.

\paragraph*{Data Availability.}
All code and data that support the findings of this article are openly available at Ref.~\cite{berceanu_multipole_symmetry_2026}.

\appendix

\section{Functional forms of the three emittance-growth channels}
\label{app:growth-laws}

The three growth channels of Sec.~\ref{sec:emittance} admit closed-form analytic descriptions whose scaling laws differ qualitatively. We collect them here for reference.

\subsection{Chromaticity}
\label{app:chromaticity}

In an unmatched drift downstream of the plasma exit, the energy-dependent betatron frequency $k_\beta = k_p/\sqrt{2\gamma}$ shears the transverse distribution and the rms normalised emittance grows according to the model of Ref.~\citep{migliorati_intrinsic_2013} [their Eq.~(5)],
\begin{equation}
\label{eq:app-chrom}
\varepsilon_n(s) = \varepsilon_{n,0}\sqrt{1 + \left(\frac{\gamma\,\sigma_\delta\,\sigma_{x'}^2\,s}{\varepsilon_{n,0}}\right)^{\!2}},
\end{equation}
linear in the drift length $s$ at large $s$. The wake remains axisymmetric throughout, placing the effect in the $m{=}0$ sector; matching the beam at the plasma exit suppresses it (Sec.~\ref{subsec:matching}).

\subsection{Hose instability}
\label{app:hose}

The bunch centroid $X_c(\xi,s)$ and dipole wake centroid $X_w(\xi,s)$ obey the coupled long-bunch system of~\citep{whittum_electron_1991,schroeder_hose_2012},
\begin{equation}
\label{eq:app-hose-coupled}
\begin{aligned}
\frac{\partial^2 X_c}{\partial s^2} + k_\beta^2 X_c &= \frac{c_r\,k_\beta^2}{2}\,X_w, \\
\frac{\partial^2 X_w}{\partial \xi^2} + k_p^2 X_w &= k_p^2 X_c,
\end{aligned}
\end{equation}
with $\xi$ the co-moving coordinate, $s$ the propagation distance, and $c_r$ an order-unity coupling coefficient. A saddle-point analysis of the coupled system yields the convective amplification
\begin{equation}
\label{eq:app-hose-growth}
X_c(\xi,s) \sim \exp\!\left[3\left(\frac{c_r\,k_\beta^2\,\xi\,s}{4\,k_p}\right)^{\!1/3}\right],
\end{equation}
the characteristic $(\xi s)^{1/3}$ law. The slice emittance inherits this growth through $\varepsilon_n \propto \langle X_c^2\rangle$ until the cavity deforms and saturates the mode. BNS damping~\citep{balakin_bns_1983} and plasma-frequency detuning~\citep{moreira_mitigation_2023} suppress it by breaking the phase-locking that closes the loop.

\subsection{Mixing resonance as a Rabi exchange}
\label{app:mixing}

When the two transverse planes are nearly degenerate, $k_{\beta,x} \approx k_{\beta,y}$, the $m{=}2$ coupling exchanges single-particle action between them in direct analogy to a two-level system. With detuning $\Delta \equiv \omega_{\beta,x} - \omega_{\beta,y}$ and coupling strength $g$, the actions obey
\begin{equation}
\label{eq:app-mix-coupled}
\begin{aligned}
\frac{dJ_x}{ds} &= -g\sin(\Delta s + \phi)\sqrt{J_x J_y}, \\
\frac{dJ_y}{ds} &= +g\sin(\Delta s + \phi)\sqrt{J_x J_y},
\end{aligned}
\end{equation}
whose single-particle solution is the Rabi formula
\begin{equation}
\label{eq:app-rabi}
J_x(s) = J_{x,0}\left[1 - \frac{g^2}{g^2+\Delta^2}\sin^2\!\left(\tfrac12\sqrt{g^2+\Delta^2}\,s\right)\right].
\end{equation}
Averaging over the bunch --- whose slices carry a spread in $\Delta$ and in initial phase $\phi$ --- damps the coherent oscillation, and on resonance ($\Delta=0$) the mean horizontal emittance relaxes monotonically toward the equipartition value
\begin{equation}
\label{eq:app-mix-ensemble}
\frac{\varepsilon_x(s)}{\varepsilon_{x,0}} \longrightarrow \tfrac12\!\left(1 + e^{-s/\tau}\right).
\end{equation}
The prefactor $g^2/(g^2+\Delta^2)$ in Eq.~\eqref{eq:app-rabi} switches off the exchange when $\Delta/g \gg 1$: detuning the transverse betatron frequencies --- the control parameter of Ref.~\citep{diederichs_resonant_2025}, Fig.~\ref{fig:diederichs-mixing} --- lifts the degeneracy and the loop cannot close.

\bibliographystyle{apsrev4-2}
\bibliography{references}

\end{document}